\documentclass[useAMS,usenatbib]{mn2e}
\usepackage{graphicx}
\usepackage{amsmath} 
\usepackage{natbib}

\title[Radio Relics]
  {Radio relics in a cosmological cluster merger simulation}
\author[Hoeft et al.]
  {Matthias Hoeft$^1$, Marcus Br\"uggen$^1$ and Gustavo Yepes$^2$ \\
  $^1$International University Bremen, Campus Ring 1, 28759 Bremen, Germany \\
  $^2$Grupo de Astrofisica, Universidad Autonoma de Madrid, Cantoblanco, 28039 Madrid, Spain}
\date{}

\pagerange{\pageref{firstpage}--\pageref{lastpage}} \pubyear{2003}

\def\LaTeX{L\kern-.36em\raise.3ex\hbox{a}\kern-.15em
    T\kern-.1667em\lower.7ex\hbox{E}\kern-.125emX}

\def\hMpc{\ifmmode{h^{-1}{\rm Mpc}}\else{$h^{-1}{\rm Mpc}$}\fi}
\def\hkpc{\ifmmode{h^{-1}{\rm kpc}}\else{$h^{-1}{\rm kpc}$}\fi}
\def\hMsun{\ifmmode{h^{-1}M_\odot}\else{$h^{-1}M_\odot$}\fi}

\begin{document}

\label{firstpage}

\maketitle

\begin{abstract}
Motivated by the discovery of a number of radio relics we investigate
the fate of fossil radio plasma during a merger of clusters of
galaxies using cosmological smoothed-particle hydrodynamics
simulations. Radio relics are extended, steep-spectrum radio sources
that do not seem to be associated with a host galaxy. One proposed
scenario whereby these relics form is through the compression of
fossil radio plasma during a merger between clusters. The ensuing
compression of the plasma can lead to a substantial increase in 
synchrotron luminosity and this appears as a radio
relic. Our simulations show that relics are most likely to be found at
the periphery of the cluster at the positions of the outgoing merger
shock waves. Relics are expected to be very rare in the centre of the
cluster where the life time of relativistic electrons is short and 
shock waves are weaker than in the cooler, peripheral regions of the
cluster. These predictions can soon be tested with upcoming
low-frequency radio telescopes.

\end{abstract}


\begin{figure*}
\includegraphics[width=0.49\textwidth,angle=-90]{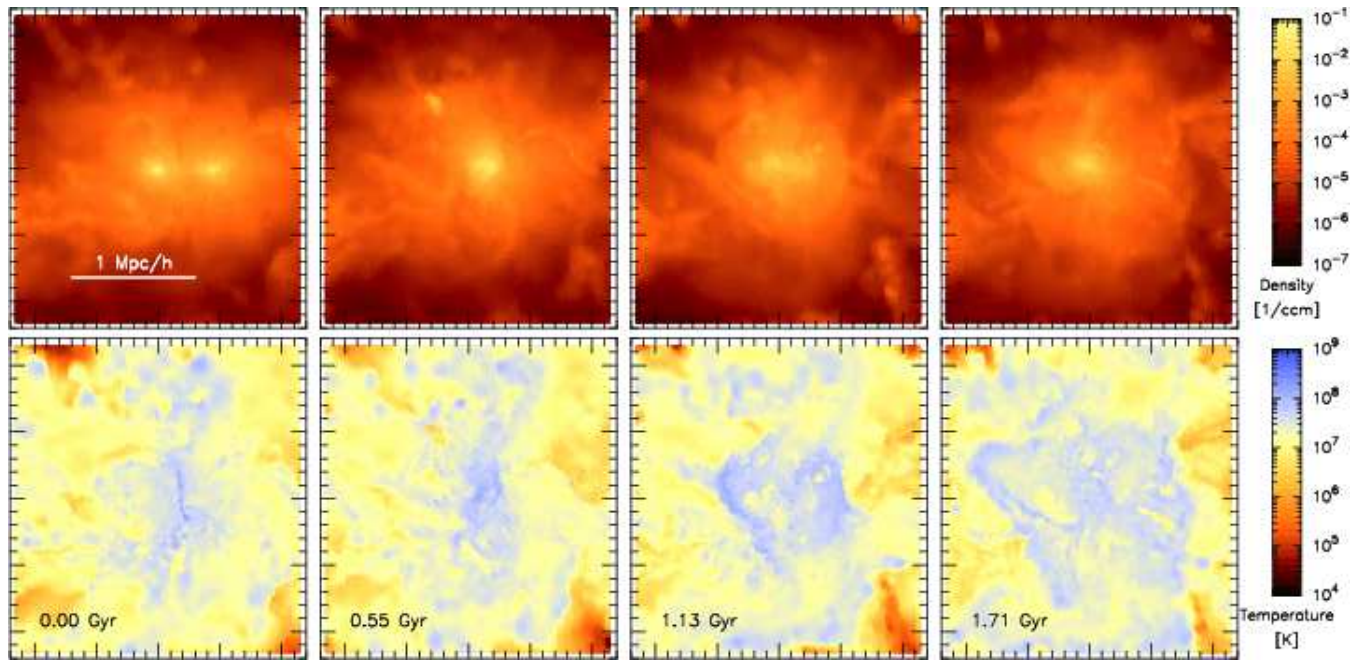}
	\caption{ The density (upper panel) and temperature (lower
		panel) distribution on a slice along a plane which contains the both
	   centres of mass of the initial two clusters.	
		The length scale is given in comoving
		$h^{-1}\,\rm{Mpc}$.  The first snapshot is taken at
		redshift $z=0.66$ and the later snapshots follow
		approximately the motion of centre of mass of the
		cluster.  The density is given in proper mass density
		divided by the proton mass.  } \label{fig-dens-temp}
\end{figure*}

\section{Introduction}

Active galactic nuclei (AGN) inject a large amount of magnetised,
relativistic plasma into the intra-cluster medium (ICM). This radio
plasma emits mainly synchrotron radiation. However, after a typical
time of 10$^8$ years the plasma has cooled radiatively such that the
remaining radio emission is difficult to detect
(e.g. \cite{jaffe:77}). The remnants of radio lobes are called `radio
ghosts' or `fossil radio plasma'. Recently, substantial evidence for
radio ghosts has come from the detection of cavities in X-ray
surface brightness maps of clusters of galaxies
\citep{boehringer:93,churazov:00,wilson:00,mcnamara:01,blanton:01}.

Diffuse, steep-spectrum radio sources with no optical identification
have been observed in a growing number of galaxy clusters. These
objects have complex morphologies that show diffuse and irregular
emission in combination with point-like sources. They are usually
subdivided into two classes: those that are located near the centre of
a cluster, e.g. in Abell 520 and Abell 2254 \citep{giovan:99}, and
those that are located in the periphery of a cluster, e.g. Abell 85,
Abell 133, Abell 3667 \citep{rottgering:97,slee:01}, denoted as `radio
halos' and `radio relics', respectively. Unlike halos, radio relics
have a filamentary morphology and show a partial polarisation of the
radio emission. This distinction, however, is not free of
contradictions. The cluster Abell 520, e.g., shows knotty radio
structures located in the centre of the cluster's X-ray emission. The
rough classification reflects the poor understanding of the origin of
the radio sources and raises the question whether halos and relics are
indeed produced by different processes.  Radio halos and relics show a
steep spectrum $\alpha \approx 1 - 1.8$ \citep{kempner:01,bacchi:03}
and the cut-off at high frequencies indicates that the electron
population has aged \citep{slee:01,kaiser:02}. For more details on
observations of diffuse cluster radio sources the reader is referred
to \citep{kempner:01, giovannini:99}. The difficulty in explaining the
radio emission lies in the lack of an evident source for the
relativistic electrons, such as an AGN. The strongest hint for the
formation of, both, halos and relics may come from the fact that both
are observed in clusters that show signs of an ongoing merger, for
instance significant substructures in the X-ray emission or the
absence of a cooling flow
\citep{bacchi:03,kempner:01,feretti:99,schuecker:99,venturi:99,roettiger:99}.
In the rest of the paper we will be primarily concerned with radio
relics because their formation is believed to be quite different from
that of radio halos.

Shock waves that are produced by a merger between clusters of galaxies
may provide the necessary acceleration of the electrons. Several
processes for the formation of radio relics have been proposed, the
two most important being: (i) in-situ diffusive shock acceleration by
the Fermi~I process \citep{ensslin:98,roettiger:99,miniati:01} and
(ii) re-acceleration of electrons by compression of existing cocoons
of radio plasma \citep{ensslin:01}. In all the aforementioned
scenarios the radio emission traces the shock-front. The fact that
radio relics resemble individual objects and do not trace the entire
shock front, may provide some indication in favour of the last
scenario.  Moreover, when a radio ghost is passed by a cluster merger
shock wave with a typical velocity of a few 1000 km/s the ghost is
compressed adiabatically and not shocked because of the much higher
sound speed within it. Therefore, diffusive shock acceleration is
unlikely to be the prime mechanism that re-energises the relativistic
electron population. However, it has been shown that the energy gained
during the adiabatic compression together with the increase in the
magnetic fields strength can cause the fossil radio cocoon to emit
radio waves again. \citet{ensslin:01} showed that the spectra thus
produced are consistent with an old electron population that has been
adiabatically compressed. With the aid of magneto-hydrodynamical
simulations \citet{ensslin:02} demonstrated that the resulting relics
possess a toroidal shape and are partially polarized - in good
agreement with observations. One prerequisite for this mechanism to be
effective is that the electron population is not older than 0.2 - 2
Gyr (En{\ss}lin \& Gopal-Krishna 2001). The time-scale depends on the
conditions in the surroundings, mainly the external pressure. In
high-pressure environments, such as in cluster cores, the synchrotron
losses are expected to be higher than in regions of lower pressure
because the magnetic fields in the pressure-confined radio plasma is
higher. This leads to a shorter synchrotron cooling time for the
fossil radio plasma and a reduced probability to flare up during the
passage of a shock.

In this paper we use numerical simulations to study the ICM during a
merger of galaxy clusters. Here we focus on the formation of radio
relics and study, in particular, whether shock waves, that are
associated with the merger, can revive the fossil plasma. The main
issues that we would like to address are (i) if radio cocoons that
have been produced before the onset of the merger can be revived by
shocks, and (ii) where the revived radio plasma is most likely found.
To produce radio relics, we apply the formalism suggested by
\citet{ensslin:01} to high-resolution simulations of a major merger.

\section{Simulations}

\label{sec-simu}

\begin{figure*}
\includegraphics[width=0.60\textwidth,angle=-90]{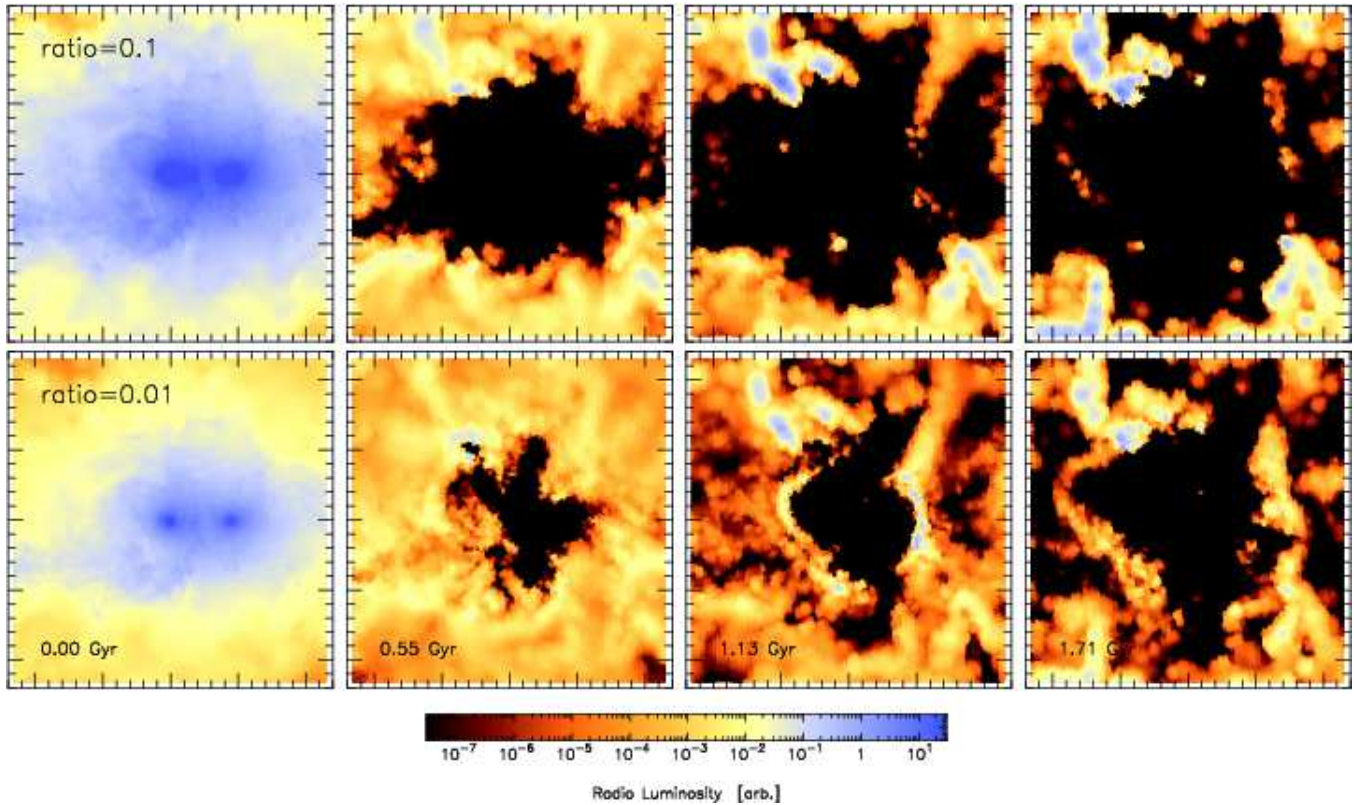}
	\caption{ 
		The radio luminosity probability at a frequency of
		100~MHz per mass of radio plasma in the same slice
		through the cluster centre as in Fig.~\ref{fig-dens-temp}. 
		The prescription for the
		evolution of the relativistic electron distribution is
		described in Sec.~\ref{sec-fate}. Here we assumed a
		pressure ratio between the magnetic and gas pressure of
		$P_B/P_{\rm{gas}}= 0.1$ and 0.01 in the upper and lower
		panels, respectively.  
		} 
	\label{fig-radio}
\end{figure*}

We performed high-resolution simulations of mergers between clusters
of galaxies using the {\sc Gadget} \citep{springel:01} code, which is
a smoothed-particle hydrodynamics (SPH) code based on a tree-scheme to
compute the gravitational interaction. We used the entropy conserving
SPH-kernel as proposed by \citet{springel:02} who found that using the
entropy as independent variable results in an improved representation
of explosion shock waves.

{\sc Gadget} provides the possibility to use particles with different
masses. This allows us to simulate a cluster merger with a high
mass-resolution following the scheme given by \citet{klypin:01}. The
simulations are set up according to the standard $\Lambda$CDM model
with $\Omega_M = 1 - \Omega_\Lambda = 0.3$, $\Omega_B = 0.039$, and
$h=0.7$, where the Hubble constant is given by
$100\:h\,\rm{km,s^{-1}\,Mpc^{-1}}$. For redshift $z=50$ the initial
particle positions and velocities are derived using a power spectrum
normalised on the $8\:h^{-1}\:\rm{Mpc}$ scale to $\sigma_8 = 0.9$.
Initially, a simulation in a computational box of
$80\:h^{-1}\,\rm{Mpc}$ with $128^3$ particles is performed. When the
simulation arrives at $z=0$, a region with a suitable cluster is
isolated. In a new simulation the mass-resolution is refined in this
region by splitting the particles up into 64 sub-particles. Thus, we
achieve a gas mass-resolution of $3.6\times10^7\:h^{-1}\,M_\odot$ in
the cluster region. Following \citet{power:03} we choose a
gravitational softening length for the refined region of
$2\,h^{-1}\,\rm{kpc}$. The final cluster is produced by a merger
between two progenitors with similar mass,
$1.7\times10^{13}\,h^{-1}\,M_\odot$ and
$1.6\times10^{13}~h^{-1}~M_\odot$, at redshift $z \approx 0.6$. The
collision processes virtually central with a relative speed of
$2000\:\rm{km \, s^{-1}}$.

We should note that effects of gas cooling, stellar feedback, thermal
conduction, and magnetic fields are not included in this
simulation. Our primary goal is to model the evolution of the merger
shock waves in the ICM within a realistic cosmological setting. In
order to calculate the radio emissivity of non-thermal plasma, we have
to make assumptions about the strength of the magnetic field. Very
little is known on the strength of the magnetic field in radio
ghosts. Measurements of the large-sclae magnetic field in clusters
indicate a field strength of few $\mu$Gauss \citep{carilli:02}. In the
centre of the cluster the field strength is higher and approximately
thermal, i.e.  $B^2/8\pi = P_B \sim P_{\rm{gas}}$,
\citep{eilek:99}. However, it seems unlikely that there is enough
small-scale turbulence or that there are enough powerful radio
galaxies to sustain a thermal magnetic field throughout the entire
cluster volume. Moreover, these measurements rely on Faraday rotation
measures, to which the dilute plasma in the radio ghosts does not
contribute significantly. Therefore, these measurements can only serve
as a very rough guide to the actual field strength in the bubbles. In
our calculations we assume that the fraction between magnetic pressure
and thermal pressure is fixed at a value which yields a mass-averaged
field strength of a few $\mu$Gauss.

\section{The fate of radio halos}

\label{sec-fate}

Relativistic electrons in radio lobes lose energy via synchrotron
radiation and inverse Compton scattering with photons of the cosmic
microwave background (CMB). In addition, adiabatic compression of the
cocoons is able to re-accelerate the electrons. On the basis of our
simulation we investigate whether the compression that the radio
plasma suffers in the course of the merger can compensate for its
radiative losses and is thus able to revive the fossil radio
plasma. In the following, we assume that each gas particle also
carries a certain amount of non-thermal radio plasma with it, whose
evolution is calculated as follows. 

The momentum of a relativistic electron in a radio lobe is altered by
synchrotron losses that are proportional to the magnetic energy
density $u_{B}$, by inverse Compton losses proportional to the CMB
field density $u_{\rm{CMB}}$ and by adiabatic compression of the
considered volume $V$
\begin{equation}
dp 
	= 
	- \frac{4}{3} \sigma_{\rm{T}} \left\{ u_{B} + u_{\rm{CMB}} \right\}
	- \frac{1}{3} \frac{1}{V}\, dV \ ,
\end{equation}
where $\sigma_{\rm{T}}$ denotes the Thomson cross
section. \cite{ensslin:01} showed that the the electron
distribution function $f(p,t)\,dp\,dV$ can be derived from the initial
distribution $f(p,t_0)$, the compression ratio
\begin{equation}
C(t) 
	= 
	V(t_0)/V(t) 
	=
	\left(
		P(t)/P(t_0)
	\right)^{1/\gamma},	
	\label{eq-compression}
\end{equation}
where the volume $V$ of the radio ghosts is adiabatically compressed
by the ambient pressure $P$ with an adiabatic exponent $\gamma$ for
magnetized plasma, and the characteristic momentum
\begin{equation}
\frac{1}{p_\ast}
	=
	\frac{4}{3} \sigma_{\rm{T}}
	\int_{t_0}^{t} \, dt' \;
	\left\{ u_{B}(t') + u_{\rm{CMB}}(t') \right\}
	\left( \frac{C(t')}{C(t)} \right)^{1/3}
	.
	\label{eq-p-ast}
\end{equation}	
If the initial distribution is a power-law $f(p,t_0) = f_0
(p/p_0)^{-\alpha}$ the spectrum at later times becomes
\begin{equation}
f(p,t)
	=
	f_0 C(t)^{(2-\alpha)/3} \,
	(p/p_0)^{-\alpha} \,
	(1-p/p_\ast)^{\alpha-2}
	.
	\label{eq-distribution}
\end{equation}
The luminosity of a fossil radio plasma depends thus on the age of the
radio plasma, the strength of the magnetic field, the CMB density and
the pressure history of the ambient gas.  To set a starting time $t_0$
of the pressure history we assume that the radio plasma has been
released at an early stage of the merger, more precisely when the two
progenitors are still separated by $0.5\,h^{-1}\,\rm{Mpc}$. Those
ghosts may result from the advection of the radio lobes by the merger,
i.e. the merger may itself have generated the radio ghosts. The
assumption that the radio plasma in the ghosts dates from the
beginning of the merger yields a conservative estimate for the later
occurrence of radio relics. As argued above, the magnetic field in the
outer part of the cluster should be sub-thermal. We assume that the
ratio $P_B/P_{\rm{gas}}$ between the magnetic and thermal pressure
is in the range of 1-10\%. Furthermore, we assume that the fossil
radio plasma moves with gas, i.e. the plasma does not feel any
buoyancy forces. This should be a tolerable assumption during the
merger itself where the motion of the radio ghosts is dominated by
advection by the ambient gas.

In our simulation we store the position, pressure and density for all
gas particles as a function of time. We assume that each particle
consists, beside the gas, of a certain amount of non-thermal radio
plasma. Thus each particle represents gas whose evolution is
calculated by the simulation and radio plasma whose evolution is
derived from the gas properties. For the starting time $t_0$ before
the merger, we assign all particles, the same power-law electron
momentum distribution. The local pressure defines the local magnetic
field strength due to the fixed ratio $P_B/P_{\rm{gas}}$ and the
compression ratio, see Eq.~(\ref{eq-compression}). We assume that the
magnetic field within the rarefied plasma bubbles is tangled on small
scales and can be approximated, together with the relativistic
particles, by a $\gamma =4/3$ equation of state. From to the recorded
pressure history of each particle alone, we can calculate the momentum
distribution of electrons in the radio plasma, see
Eq.~(\ref{eq-distribution}). The luminosity of the radio plasma is
computed for an observing frequency of $\nu = 100\ ~\rm{MHz}$ and an
initial spectral index of $\alpha = 2.5$ using the standard
integration kernel for synchrotron radiation \citep{rybicki:79}.

\begin{figure}
\includegraphics[width=0.36\textwidth,angle=-90]{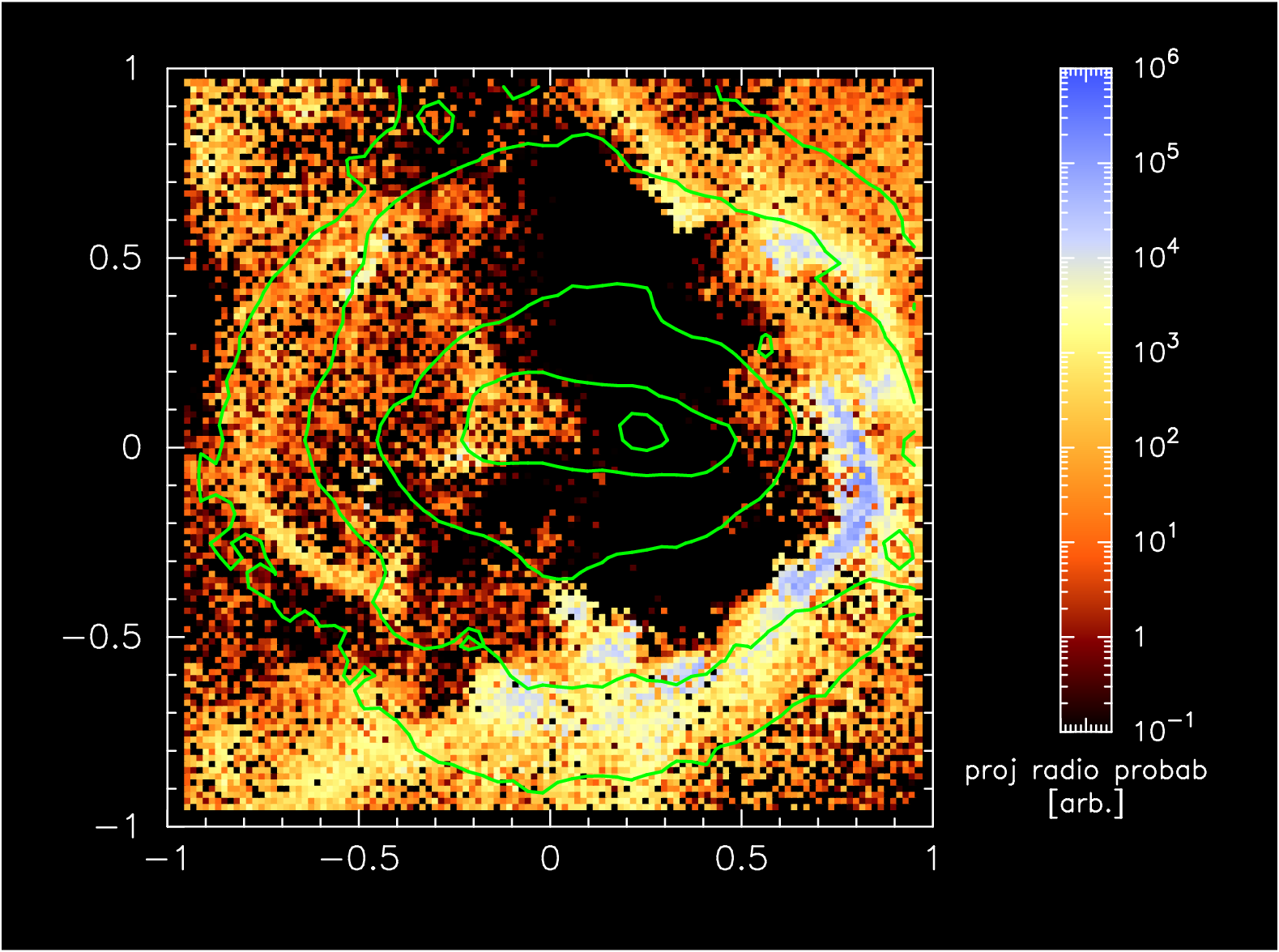}
	\caption{ The projected 'potential' radio luminosities for
1.13~Gyr old radio plasma, where $P_B/P_{\rm{gas}} = 0.01$. The radio
plasma is identically to the gas distributed, see
Sec.~\ref{sec-discuss}.  For comparison the bolometric surface X-ray
luminosity, $ L_X = 1.2\times 10^{-24}\rm{erg s^{-1}} \, m_{\rm{gas}}
/ (\mu m_p) \, \sum \, \rho_i/ (\mu m_p) \, ( kT_i / \rm{keV} )^{1/2}$
\citep{eke:98}, is given.  Contours are at $10^{41}$, $10^{42}$,
$10^{43}$, $10^{44}$ and
$10^{45}\,\rm{erg}\,\rm{s}^{-1}\,h^3\,\rm{Mpc}^{-3}$.  The total
bolometric X-ray of the cluster is
$2\times10^{44}\,\rm{erg}\,\rm{s}^{-1}$ and the emission-weighted
temperature is $3\,\rm{keV}$.  } \label{fig-projected}
\end{figure}

\section{Results and Discussion}

\label{sec-discuss}

The merger produces shock waves that propagate in both directions
along the line that connects the centres of the initial
clusters. While the shock generates only a small jump in density, the
temperature in the shocked regions is about one order of magnitude
above that in regions in front of the shock (see
Fig.~\ref{fig-dens-temp}). The effects of the merger on the radio
plasma can be seen in Fig.~\ref{fig-radio} which shows the radio
luminosity in a slice through the cluster. The slice was made along
the plane which contains the both centres-of-mass of the initial two
clusters. The most remarkable structure is the prominent ring-like
feature with a diameter of about 1~Mpc (see Fig.~\ref{fig-radio})
1.13~Gyr after the release of the radio plasma.  This structure
corresponds to the outgoing shock waves seen in the temperature (see
Fig.~\ref{fig-dens-temp} lower panel). One can clearly see the flaring
of radio plasma at the two outgoing merger shock waves. It is apparent
that after about 1~Gyr the merger shock waves can still revive the
fossil radio plasma. In contrast, it is striking that the cluster
centre is virtually void of any luminous sources. Prerequisite for a
reanimation of the plasma is a sufficiently low magnetic field
strength. Only if the strength is as low $P_B/P_{\rm{gas}} \sim 1\%$
revived structures can be seen. Higher magnetic field strengths result
in too fast an ageing of the plasma, such that the shock waves cannot
revive the plasma. Furthermore, in the early stage of the merger, when
the shock waves pass through the centre of the cluster after about
0.5~Gyr, the even younger plasma does not flare up significantly.

The much higher occurrence of radio relics at peripheral locations in
the cluster is a result of two factors: In the centre of the cluster,
i.e. within a region a few hundred kpc in diameter, the radio plasma
ages much faster because the pressure is much higher and, thus, also the
(assumed) magnetic field. This causes considerably higher radiation
losses. In the inner region of the cluster the luminosity of the
plasma dies down after $\approx 0.5\:\rm{Gyr}$, whereas in the
periphery the luminosity decreases much more slowly. The second reason
is related to the shock compression. As the shocks sweep through the
cluster, their strength varies with the ICM sound speed. The shock
waves are relatively weak in the hotter cluster centre but steepen
when they pass the cooler, outer regions of the cluster. Therefore,
the compression factor of the shock increases and, thus, the ability to
revive radio ghosts. Under the condition that the ratio
$P_B/P_{\rm{gas}}$ is as low as $\approx 1 \%$, the two shock waves
can lighten up fossil radio plasma in the outer regions of the cluster
which is about 1~Gyr old. In stronger fields, synchrotron losses are
too severe and in weaker fields, higher energy electrons are required
to emit at the observing frequency, which have more severe
inverse-Compton losses. For $P_B/P_{\rm{gas}}\approx 1 \%$ the two
losses roughly match. We should point out again that
Fig.~\ref{fig-radio} and \ref{fig-projected} are luminosity
probability maps and not actual maps. Therefore, we do not expect
radio relics over the entire shock surface.

We find additional bright spots in the periphery of the cluster which
are not related to the shock fronts. These spots exist even if
$P_B/P_{\rm{gas}}$ is as high as $\approx 10\%$. Following the
evolution of merger, one can notice that these spots appear in regions
where gas from the periphery is flowing into the cluster. Since the
radio plasma from regions that lie further out has aged less than more
centrally located plasma, a merely moderate compression may cause a
noticeable increase in the luminosity. This provides an additional
mechanism to generate radio relics, which again only takes place in
clusters that are not in equilibrium. Both mechanisms, gas infall and
merger shock compression, strongly favour radio ghosts that are
located at least at a distances of a few hundred kpc from the cluster
centre.

We now compute the radio surface brightness. For this purpose we
convolve the luminosity of radio plasma with a hypothetical
distribution of radio plasma. Here, we assume that the distribution of
plasma follows the distribution of gas. This approach disregards the
fact that there are -- probably -- only few individual plasma object
in a cluster, but we interpret the distribution of the entire cluster
as a probability that plasma is located at a certain position. We can
now infer the radio luminosity of the cluster. Given that the
distribution of plasma reflects the spatial probability distribution
of radio ghosts, the luminosity corresponds to the probability to find
radio relics at a certain position. We find that by far most of the
luminosity projected onto a plane perpendicular to the shock fronts
comes from the shock regions (see Fig.~\ref{fig-projected}). Even the
projection leaves the central region of the cluster virtually free of
emission. This result indicates, that if the distribution of radio
ghosts follows the distribution of the gas, i.e. if ghosts are most
likely found in the centre, relics are expected to be observed almost
exclusively at the location of the shock fronts.

Upcoming radio telescopes, such as GMRT, LOFAR and ALMA, will identify
more radio relics and will measure their spatial distribution. They
will be able to verify whether relics are predominantly found along
outgoing merger shock waves at distances of a few hundred kpc from the
centre of the cluster \citep{ensslin:02}. Thus, in the near future,
observations of a representative sample of radio relics will put our
model to test.

In summary, on the basis of cosmological high-resolution simulations
of a galaxy cluster merger with two progenitors of the mass $\sim
1.6\times10^{13}\:h^{-1}\,M_\odot$ we have been able to calculate,
both, the fading of non-thermal radio plasma and its re-energisation
by shock waves. It was found that cluster-wide shock fronts are
capable of reviving $\sim 1\:\rm{Gyr}$ old radio ghosts if the
ratio $P_B/P_{\rm{gas}}$ is as low as $1\%$. Magnetic field strengths
of this order are not unlikely in the cluster periphery. Finally, the
vast majority of radio relics are expected to be located at typical
distances of a few hundred kpc from the cluster centre.  \\

{\sc Acknowledgement} We thank Stefan Gottl\"ober for providing the
initial conditions for the simulation and Volker Springel for
providing the version of {\sc Gadget} with the entropy kernel.  MB
thanks Torsten En{\ss}lin for helpful discussions and valuable
comments.  The simulation were performed at CIEMAT (Spain) and at the
Konrad-Zuse-Zentrum f\"ur Informationstechnik Berlin.  GY and MH
thanks for financial support from the Acciones Integradas
Hispano-Alemanas.

\newcommand{\aap  }{A\&A}
\newcommand{\araa }{ARA\&A}
\newcommand{\apj  }{ApJ}
\newcommand{\apjl }{ApJL}
\newcommand{\aj   }{AJ}
\newcommand{\mnras}{MNRAS}

\bibliography{radio}
\bibliographystyle{apj}

\end{document}